\documentclass[twoside,fleqn]{article}
\usepackage{espcrc2}
\newcommand{\be}{\begin{equation}}
\newcommand{\ee}{\end{equation}}
\newcommand{\bea}{\begin{eqnarray}}
\newcommand{\eea}{\end{eqnarray}}
\newcommand{\pb}{\(\pi\beta\;\)}
\newcommand{\nn}{\nonumber}
\newcommand{\hs}{\hspace*}
\title{
\vspace{-1.0cm}
{\sf \small \rightline{\large{PSI-PR-02-09}}}
\bigskip
Electromagnetic radiative corrections to pionic beta decay
\(\pi^+\rightarrow \pi^0 e^+\nu_e\)\thanks{
Talk presented at QCD 02, Montpellier, 2-9 July 2002. This article is based 
on work done together with V. Cirigliano, M. Knecht and H. Neufeld.}}

\author{H. Pichl
\vspace*{0.15cm}\newline
Paul Scherrer Institut, Theory Group, CH-5232 Villigen PSI, 
Switzerland}
\begin{document}
\begin{abstract}
Pionic beta decay \(\pi^+\rightarrow \pi^0 e^+\nu_e\) is analyzed in chiral 
perturbation theory with virtual photons and leptons. All 
electromagnetic corrections up to \(\mathcal{O}\)\((e^2 p^2)\) are
taken into account. Theoretical results are confronted with preliminary data 
from a PSI measurement and a value for the CKM matrix element \(|V_{ud}|\) is
given. Although the precision is presently still below the 
one of existing determinations of \(|V_{ud}|\), an analysis of pionic beta 
decay, based on a systematic treatment within a low-energy effective field 
theory, may become a useful alternative.
\end{abstract}
\maketitle
\section{Introduction}

One of the cornerstones of the Standard Model is the CKM matrix \(V.\) Its 
matrix elements \(V_{ij}\) determine the strengths of transitions between 
quark flavours \(i\) and \(j\). The \(V_{ij}\) are therefore of eminent 
importance, and their precise determination is a major task of today's 
particle physics.

Unitarity implies a series of relations among the matrix elements. In 
particular, one expects the first-row CKM matrix elements to satisfy
\be
\label{US}
|V_{ud}|^2+|V_{us}|^2+|V_{ub}|^2=1,
\ee
where mixing among the first-generation quarks \(u\) and \(d\) is 
strongest, and mixing between the first and third generation is very small and
can be neglected at the present precision \cite{PDG}. 

The strength of an \(s\) to \(u\) transition is given by \(V_{us}\) 
which is measured in \(K_{\ell 3}\) decays. Taking into account a recent 
analysis of radiative corrections in \(K_{\ell 3}\) decays in chiral 
perturbation theory \cite{CIR}, 
the Particle Data Group (PDG) leaves the central value of \(|V_{us}|\) 
unchanged but increases the error and gives finally 
\(|V_{us}|=0.2196\pm 0.0026\) \cite{PDG}. 

The most precise up-to-date knowledge of \(|V_{ud}|\) comes either from 
analyses of nuclear beta decays, in particular super-allowed Fermi 
decays, or (not as precise) from neutron decay. 

From beta decays of nuclei one finds \(|V_{ud}|=0.9740\pm0.0005\), where the 
uncertainty depends on structure-dependent radiative corrections \cite{PDG}. 
Interactions with the nuclear medium could give the quarks different 
effective masses, which might lead to an enhanced value of \(|V_{ud}|\). 
The PDG compensates for this possibility by doubling the aforementioned 
error \cite{PDG}.

An extraction of \(|V_{ud}|\) from neutron decay benefits on one hand from 
fewer theoretical problems, but suffers at the same time from the fact that 
not only the neutron lifetime but also the ratio of axial-vector and vector 
couplings \(g_A/g_V\) comes into play. Averaging over neutron decay 
experiments (see \cite{PDG} and references therein), the PDG gives 
\(|V_{ud}|=0.9725\pm0.0013\), and a final combined value of 
\(|V_{ud}|=0.9734\pm0.0008\) with a remarkably small error is found 
\cite{PDG}. 

With these values one finds a \(2.2\sigma\) deviation from unitarity in
(\ref{US}): \(|V_{ud}|^2+|V_{us}|^2+|V_{ub}|^2-1 = -0.0042\pm 0.0019\). 
Adopting a conservative line of reasoning, one may speculate if the given 
error of \(|V_{ud}|\) might be too optimistic, and look for a cleaner 
theoretical alternative to determine \(|V_{ud}|\). 

Pionic beta (\(\pi\beta\)) decay \(\pi^+\rightarrow \pi^0 e^+\nu_e\) combines
the advantages of the traditional approaches, i.e., it is a pure vector 
transition without any complications due to nuclear structure. However, 
practical utilization of \pb decay is severely complicated by its branching 
ratio of \(\rm BR_{\pi\beta}\sim 1\times 10^{-8}\) 
\cite{McF,PIBETA}. 

In spite of the rareness of the decay, a measurement of \(\rm BR_{\pi\beta}\) 
is being performed at PSI, aiming to achieve a precision of \(0.5\%\) in a 
first stage \cite{PIBETA}. Of course, this first accuracy will not be enough 
to compete with nuclear or neutron decays, but a third independent 
determination of \(|V_{ud}|\) may certainly be useful. 
\newpage
At this level of accuracy radiative corrections have to be included in the 
theoretical analysis, and we employ chiral perturbation theory as the ideal 
framework for low-energy hadron physics\footnote[1]{Jaus also \cite{Jaus} 
addressed the problem of rad. corrections to \pb decay in a constituent quark 
model.}. This article is based on \cite{CIR2} which in turn is based on 
recent work on \(K_{\ell 3}\) decays \cite{CIR}.    

\section{Kinematics and decay amplitude}

Neglecting radiative corrections for the moment, the decay amplitude of 
\pb decay
\be
\pi^+(p_+)\rightarrow\pi^0(p_0)\;e^+(p_e)\;\nu_e(p_{\nu})
\ee
is written in terms of two form factors \(f^{(0)}_{\pm}\):
\be
\mathcal{M}=G_FV^{\ast}_{ud}\:l_{\mu}\left[(p_+^{\mu}+p_0^{\mu})f^{(0)}_+ + 
(p_+^{\mu}-p_0^{\mu})f^{(0)}_-\right]\nn\!\!,
\ee
where \(l_{\mu}=\bar{u}(p_{\nu})\gamma_{\mu}(1-\gamma_5)v(p_e)\) is the lepton
current. \(G_F\) is Fermi's coupling constant. 

At leading order, \(f^{(0)}_+=1\) and \(f^{(0)}_-=0\). At next-to-leading 
order, the form factors develop a dependence on the variable
\(t=(p_+-p_0)^2\). 

\(f^{(0)}_-\) turns out to be proportional to the difference of squared pion 
masses \((m_{\pi^+}^2-m_{\pi^0}^2)\). Additionally, \(f^{(0)}_-\) terms in 
the squared decay amplitude come always along with a factor of 
\(r_e=m_e^2/m_{\pi^+}^2\simeq 1.35\times 10^{-5}\). These observations justify 
to neglect \(f^{(0)}_-\) in the amplitude. 

The decay rate \(\Gamma_{\pi\beta}\) for \pb decay,
\be
\label{DW}
\Gamma_{\pi\beta}=\int_
{\mathcal{D}} dy\,dz \;\rho^{(0)}(y,z)\;, 
\ee
is expressed in terms of a phase space density \(\rho^{(0)}\) which depends 
on two variables \(y\) and \(z\) defined in the c.m.s. by
\be
y=\frac{2E_e}{m_{\pi^+}}\;\;\;\mbox{and}\;\;\; z=\frac{2E_{\pi^0}}
{m_{\pi^+}}\,.
\ee
\(\mathcal{D}\) denotes the kinematically allowed area of \((y,z)\). The 
phase space density itself is given by 
\be
\label{PSD}
\rho^{(0)}(y,z)=\mathcal{N}\times|f^{(0)}_+(t)|^2A^{(0)}(y,z)\;,
\ee
where 
\(A^{(0)}(y,z)\) is a kinematical density
\bea
\label{KD}
&&\hs{-0.6cm}A^{(0)}(y,z)=\left[ 4\left(-1+z+2y-yz-y^2\right)-4r_0\right.\nn\\
&&\hspace*{0.95cm}+\left. r_e \left(4y+3z-3+r_0-r_e \right) \right]\,,
\eea
and \(\mathcal{N}\) in (\ref{PSD}) and \(r_0\) in (\ref{KD}) are defined by
\be
\mathcal{N}=\frac{|V_{ud}|^2\,G_F^2 \,m_{\pi^+}^5}{64\pi^3}\,,\;\;\;\;\;\;
r_0=m_{\pi^0}^2/m_{\pi^+}^2\,. 
\ee 

At \(\mathcal{O}\)\((p^4)\), meson loops and counterterms from the chiral
Lagrangian \(\mathcal{L}_{4}\) \cite{GL1} generate the aforementioned 
\(t\)-dependence\footnote[2]{\(\mathcal{O}(e^2p^0)\) corrections are 
absorbed in the meson masses.}: \(f^{(0)}_+(t)=1+\delta f^{(0)}_+(t)\),
\bea
\delta f^{(0)}_+(t)&=&\frac{1}{F^2}\left[2tL^r_9(\mu)+2h^r_{\pi^+\pi^0}(t,\mu)
\right.\nn\\
&&\hspace*{0.55cm}\left.+h^r_{K^+K^0}(t,\mu)\right].
\eea
The functions \(h^r_{PQ}(t,\mu)\) comprise one-loop corrections from the 
particles \(P\) and \(Q\) \cite{GL1}. Both the renormalized low-energy 
coupling (LEC) \(L^r_9(\mu)\) and \(h^r_{PQ}(t,\mu)\) depend individually on 
the ren. scale, but \(\delta f^{(0)}_+(t)\) is independent of \(\mu\).

\section{Radiative corrections}

Radiative corrections manifest themselves in virtual photon exchange, new 
electromagnetic counterterms, and in the emission of real photons. 

Photon exchange generates one-loop diagrams as a result of which a new 
dynamical variable \(u=(p_+-p_e)^2\) and infrared divergences appear in 
the amplitude \(\mathcal{M}\). Counterterms from 
Lagrangians \(\mathcal{L}_{e^2p^2}\) and \(\mathcal{L}_{\rm lept}\) 
\cite{Urech,KNRT} contribute three new LECs \(X_1\), \(X^r_6(\mu)\), and 
\(K^r_{12}(\mu)\). 

Virtual radiative corrections induce a change of the form factor from 
\(f^{(0)}_+(t)\) to \(f_+(t,u,\lambda)=f^{(0)}_+(t)+\delta f^{\rm em}_+
(u,\lambda)\), where \(\lambda\) plays the role of a fictitious photon mass
to regularize infrared divergences. 

Only showing local contributions explicitly and omitting remaining one-loop
contributions, the radiative corrections of \(\mathcal{O}(\alpha)\) read
\bea
&&\hs{-0.7cm}\delta f^{\rm em}_+(u,\lambda)=\delta f^{\rm em}_{+\ell}
(u,\lambda)+\delta f^{\rm em}_{+\:\rm local}\nn\,,
\eea
\bea
\label{DRC}
&&\hs{-0.7cm}\delta f^{\rm em}_{+\:\rm local}=4\pi\alpha\left\{-\frac{2}{3}
X_1-\frac{1}{2}X_6^r(\mu)+2K^r_{12}(\mu)\right.\nn\\
&&\hs{0.6cm}\left.-\frac{1}{32\pi^2}\left [3+\ln{\frac{m_e^2}{m_{\pi^+}^2}}
+3\ln{\frac{m_{\pi^+}^2}{\mu^2}}\right ]\right\}.
\eea
As before, \(\delta f^{\rm em}_+(u,\lambda)\) is independent of the 
renormalization scale\footnote[3]{The omitted one-loop part 
\(\delta f^{\rm em}_{+\ell}(u,\lambda)\) containing in particular the infrared 
divergences can be found in \cite{CIR2}.}.   

As it was shown in \cite{CIR,CIR2}, it is useful to rearrange the complete 
form factor in the following way:
\be
f_+(t,u,\lambda)=F_+(t)\times\left[1+\delta f^{\rm em}_{+\ell}(u,\lambda)
\right]\,,
\ee
by factoring-out the \(t\)-dependence in \(F_+(t):=f^{(0)}_+(t)+\delta 
f^{\rm em}_{+\:\rm local}\) and isolating both infrared divergences and 
\(u\)-dependence. 
This allows us to write the rate in the presence of virtual radiative 
corrections in combination with real photon emission formally in exactly the 
same way as in (\ref{DW}), (6), i.e., in terms of a generalized form 
factor \(F_+(t)\) and a generalized infrared finite kinematical density. 

The treatment of real photon emission and cancellation of infrared divergences
was discussed in detail in \cite{CIR,CIR2}. Working at order \(\alpha\), it
is sufficient to consider the emission of only one real photon. We adopt a 
scheme proposed by Ginsberg \cite{Gi1} and integrate over the entire
phase space of undetected particles, i.e., the photon and the neutrino, but 
restrict the \((y,z)\) space to the non-radiative three-particle decay case. 
Furthermore, a new variable \(x=(p_{\nu}+p_{\gamma})^2\) is defined and 
integrated over.

The phase space density for the radiative decay 
\(\pi^+\rightarrow \pi^0 e^+\nu_e\gamma\) is given by
\bea
\label{DWG}
&&\hs{-0.6cm}\rho^{\gamma}(y,z) =  \frac{m_{\pi^+}}{2^{13}\pi^6} 
\int\limits_{\lambda^2}^{x_{\rm max}}dx\int\frac{d^3 p_\nu}{E_\nu} 
\frac{d^3 p_\gamma}{E_\gamma}\nn\\ 
&&\hs{-0.3cm}\times\,\delta^{(4)}(p_+-p_0-p_e-p_\nu-p_\gamma)\sum_{\rm pol}
|{\cal M}^{\gamma}|^2\,,
\eea 
where \(\mathcal{M}^{\gamma}\) denotes the leading-order amplitude of the 
radiative decay, respectively. Upon combining the phase space densities of 
the mother process and the associated radiative decay, one arrives at an 
infrared finite phase space density 
\be
\rho(y,z)=\rho^{(0)}(y,z)+\rho^{\gamma}(y,z)
\ee 
which we may write, as already anticipated, as 
\be
\rho(y,z)=\mathcal{N}\times\left|F_+(t)\right|^2\times A(y,z)\,,
\ee
in terms of a generalized kinematical density \(A(y,z)\) and the form factor 
\(F_+(t)\). Apart from electromagnetic local contributions, radiative 
corrections from virtual photon exchange (\ref{DRC}) and real photon emission 
(\ref{DWG}) are entirely included in \(A(y,z)\) \cite{CIR,CIR2}.

\section{Results and conclusions}

Before presenting our results, a few comments seem advisable. In general, 
low-energy couplings of chiral Lagrangians parametrize short-distance 
physics that is not explicitly dealt with in the effective theory. 

According to \cite{SIR,MS}, it turns out that all semileptonic charged 
current amplitudes are affected by universal short-distance physics 
corrections when expressed in terms of the muon decay constant. Relating 
therefore \(G_F\) with the muon lifetime, it is possible to trace back some of 
the high-energy origin of the LEC \(X_6\) to these universal corrections 
\cite{CIR}. Hence, we may write \(X^r_6(\mu)=X_6^{\rm SD}+\tilde{X}^r_6(\mu)\),
and define \(e^2X^{\rm SD}_6=1-\mathcal{S}_{\rm ew}(m_{\rho},m_Z)\) to 
contain the short-distance enhancement factor 
\(S_{\rm ew}(m_{\rho},m_Z)=1.0232\) from \cite{CIR,MS}. The remaining part 
\(\tilde{X}^r_6(\mu)\) is now expected to be of the order of \(\sim 10^{-3}\). 

Extracting the short-distance physics from \(\rho(y,z)\) and expanding until 
\(\mathcal{O}(t)\), we may write the infrared safe inclusive decay rate as 
\be
\label{IDW}
\Gamma_{\pi\beta(\gamma)}=\mathcal{N}S_{\rm ew}(m_{\rho},m_Z)\times|F_+(0)|^2
I(\lambda_+).
\ee
The linear approximation works extremely well, since \(m_e^2\leq t \leq 
(m_{\pi^+}-m_{\pi^0})^2\simeq 21.1\mbox{ MeV}^2\). 
\be
I(\lambda_+)=\int_{\mathcal{D}} dy\,dz A(y,z)\left[1+\frac{t}{m_{\pi^+}^2}
\lambda_+\right]^2
\ee
contains the slope parameter \(\lambda_+\) from the expansion in \(t\).

To estimate \(F_+(0)\), we use the 'classical' value of 
\(L^r_9(m_{\rho})=(6.9\pm0.7)\times 10^{-3}\), quoted in \cite{GL1}, and take 
Moussallam's estimate \cite{MOU} of \(K^r_{12}(m_{\rho})=(-4.0\pm0.5)\times 
10^{-3}\). 

Since essentially nothing\footnote[4]{In a combined analysis of 
\(K^+_{\ell 3}\) and \(K^0_{\ell 3}\) decays it is possible to extract a 
value for \(X_1\) \cite{VIN}.} is known about the \(X_i\), we resort to 
naive dimensional analysis and use \(|X_1|,\;|\tilde{X}^r_6(m_{\rho})
|\leq 6.3\times 10^{-3}\). 

Finally, \(F_+(0)\) is found to be \(1.0046\pm 0.0005\). The error is 
completely due to uncertainties in the electromagnetic LECs and is 
extremely small. The slope parameter's uncertainty comes from the error of 
\(L^r_9(m_{\rho})\), and we obtain \(\lambda_+=(0.037\pm0.003)\). 

The phase space factor \(I(\lambda_+)\) is calculated to be \(I(\lambda_+)
\simeq7.3832\times 10^{-8}\) with a tiny error.

We arrive at the following results. Radiative corrections turn 
out to be at the \(\le 1\%\) level. The corrected phase space \(I(\lambda_+)\)
enhances the rate by \(\sim 0.1\%\), the form factor \(F_+(0)\) by 
\(\sim 0.9\%\).  

Using the PDG's value of \(|V_{ud}|=0.9734\pm0.0008\), we arrive at a 
\(\pi\beta\) branching ratio of
\be 
\rm BR_{\pi\beta}=(1.0376\pm0.003)\times 10^{-8}.
\ee   

Turning the procedure upside down, we use the \pb branching ratio
together with our calculation of radiative corrections to extract 
\be
|V_{ud}|=9600.8\times\frac{\sqrt{\rm BR_{\pi\beta(\gamma)}}}{
|F_+(0)|}.
\ee
The error assigned to our value of \(|V_{ud}|\) is
\be
\Delta|V_{ud}|=|V_{ud}|\times\left[\pm\frac{\Delta F_+(0)}{F_+(0)}\pm\frac{1}
{2}\frac{\Delta \rm BR_{\pi\beta (\gamma)}}{\rm BR_{\pi\beta (\gamma)}}\right]
\ee

For the PDG's value of the branching ratio 
\(\rm BR_{\rm PDG}=(1.025\pm0.034)\times 10^{-8}\), we get
\be
|V_{ud}|_{\rm th\,1}=0.9675\pm0.0005(\mbox{th})\pm0.017(\mbox{exp}).
\ee
Obviously, the theory is very well under control and it is the experiment 
that limits the precision. 

The situation is much improved if one uses new PSI data on 
\pb decay \cite{PIBETA}. With the recent, however still preliminary, 
branching ratio of \(\rm BR_{\rm PSI}=(1.044\pm0.016)\times 10^{-8}\), we find
\be
\label{estimate}
|V_{ud}|_{\rm th\,2}=0.9765\pm0.0005(\mbox{th})\pm0.0075(\mbox{exp}).
\ee 

From the theorist's point of view, we wish to emphasize that the inclusion of 
radiative corrections is extremely straightforward and particularly simple. 
The remarkably small theoretical error can practically entirely be 
attributed to the low-energy couplings \(X_1\) and 
\(\tilde{X}^r_6(m_{\rho})\). 

Moreover, the approach is completely equivalent to the one employed in 
\(K_{\ell 3}\) decays \cite{CIR}, which puts both processes on an equally 
sound footing. Note finally that the central value of \(|V_{ud}|\) in 
(\ref{estimate}) matches the unitarity constraint (\ref{US}) very well, but 
the error is, of course, much too large to allow for more stringent final 
conclusions.       

It is fair to say that the present experimental precision of \pb decay cannot 
yet compete with nuclear beta decays, and even if the final precision will 
drastically increase it remains rather unlikely to reach a competitive 
accuracy. Nevertheless, \pb decay could become a valuable alternative source 
for the determination of \(|V_{ud}|\), in particular if previous error 
estimates turned out to have been too optimistic.
%
% List of references
%

\end{document}